# Two-dimensional van der Waals Heterostructures for Synergistically Improved Surface Enhanced Raman Spectroscopy

*Qiran Cai,[1][†] Wei Gan,[1][†] Alexey Falin,[1,2] Kenji Watanabe,[3] Takashi Taniguchi,[3] Jincheng Zhuang,[4] Weichang Hao,[4] Shaoming Huang,[5] Tao Tao,[2,6] Ying Chen,[1] Lu Hua Li[1]\**

1. Institute for Frontier Materials, Deakin University, Waurn Ponds Campus, Waurn Ponds, VIC 3216, Australia.

2. Dongguan South China Design Innovation Institute, Dongguan 523808, China

3. National Institute for Materials Science, Namiki 1-1, Tsukuba, Ibaraki 305-0044, Japan.

4. BUAA-UOW Joint Research Centre and School of Physics, Beihang University, Beijing 100191, China

5. Guangdong Key Laboratory of Low Dimensional Materials & Energy Storage Devices, School of Materials and Energy, Guangdong University of Technology, Guangzhou 510006, China

6. Guangdong Provincial Key Laboratory of Functional Soft Condensed Matter, School of Materials and Energy, Guangdong University of Technology, Guangzhou 510006, China

† These authors contributed equally.

ABSTRACT

Surface enhanced Raman spectroscopy (SERS) is a precise and non-invasive analytical technique that is widely used in chemical analysis, environmental protection, food processing, pharmaceutics, and diagnostic biology. However, it is still a challenge to produce highly sensitive and reusable SERS substrates with minimum fluorescence background. In this work, we propose the use of van der Waals heterostructures of two-dimensional materials (2D materials) to cover plasmonic metal nanoparticles to solve this challenge. The heterostructures of atomically thin boron nitride (BN) and graphene provide synergistic effects: (1) electrons could tunnel through the atomically thin BN, allowing the charge transfer between graphene and probe molecules to suppress fluorescence background; (2) the SERS sensitivity is enhanced by graphene *via* chemical enhancement mechanism (CM) in addition to electromagnetic field mechanism (EM); (3) the atomically thin BN protects the underlying graphene and Ag nanoparticles from oxidation during heating for regeneration at 360 °C in the air so that the SERS substrates could be reused. These advances will facilitate wider applications of SERS, especially on the detection of fluorescent molecules with higher sensitivity.

INTRODUCTION

Surface enhanced Raman spectroscopy (SERS) is the only non-invasive analytical technique capable of providing information on chemical bonds of molecules at extremely low concentrations (down to a single molecule), and hence the technique has been widely used in chemical analysis, environmental protection, detection of illicit drugs, art preservation, forensic science, biological diagnosis, and medical industry etc.[1-4] The signal enhancement is generally achieved *via* two mechanisms: electromagnetic field mechanism (EM) and chemical/charge transfer mechanism (CM).[5-7] EM is associated with local electromagnetic fields enhanced by localized surface plasmon resonance (SPR) with an enhancement factor (EF) up to $10^8$.[8] Noble metal (such as Au, Ag and Cu) nanoparticles are commonly used in EM for maximal enhancements due to their plasmon resonance frequencies in the visible and near-infrared range.[9-11] CM is based on charge transfer between probe molecules and substrates, increasing molecular polarizability and hence Raman cross-section.[12]

In spite of tremendous efforts in the field, it is still a challenge to achieve highly sensitive, reproducible, homogeneous, reusable SERS substrates with minimum fluorescence background. Fluorescence background is a serious problem in SERS, especially when visible light is used for excitation. With its cross-section about 6 orders larger than that of Raman scattering,[13] fluorescence as a strong background signals increases noise levels and disguises weak Raman signals, greatly affecting its accuracy and sensitivity. Although the cross-section of Raman can be increased by $10-10^2$ *via* CM and $10^3-10^8$ *via* EM, fluorescence is enhanced by $10-10^3$ at the same time. That is, fluorescence is still at least 10 times stronger than Raman signals. Electrically conductive substrates can quench fluorescence background by conducting excited electrons away. The use of excitation light source in the near-infrared region can also alleviate the problem, but some important wavelength-dependent Raman information is lost.[14] On the other hand, the reusability of SERS substrates is important to its wider applications because of the high costs of noble metals and fabrication processes involved. Heating at above 350 ºC in the air or oxygen-containing gas has been found the only effective regeneration method to remove probe molecules on SERS substrates for reuse.[15-16]

Two-dimensional (2D) materials, such as graphene, atomically thin boron nitride (BN) and transition metal dichalcogenides (TMDs) [12,17-19] provide new possibilities for SERS due to their unique structure and properties. When covering plasmonic nanoparticles, graphene greatly improves the signal to noise ratio of Raman by preserving the SPR and effectively quenching fluorescence.[13,20] Graphene also improves the stability of Ag particles in ambient due to its high impermeability.[17] However, this protection is only temporary, and in the long term graphene accelerates metal oxidation due to galvanic corrosion.[21-22] Another drawback of graphene is its low thermal stability:[18] it starts to react with oxygen at ~250 ºC.[23] As a result, it cannot protect Ag particles from oxidation at higher temperatures for the regeneration and reuse of SERS substrates.

Atomically thin BN has also been explored for its application in SERS. It has a very similar honeycomb structure as graphene and possesses many excellent chemical and physical properties.[11,16,18-19,24-26] For example, atomically thin BN showed much stronger absorption capability than bulk hBN towards aromatic molecules due to conformational change, which can increase SERS sensitivity.[27] Atomically thin BN-veiled SERS substrates showed an up-to two orders enhancement in Raman sensitivity compared with conventional SERS substrates.[18] More importantly, BN made SERS substrates reusable by heating in air at ~360 ºC to remove adsorbed probe molecules, and ~90% sensitivity remained after 30 times of reuse.[18] In addition, atomically thin BN has long term protection on metals thanks to its electric insulation.[18,28] However, BN could not quench fluorescence, deterring a higher detecting sensitivity for fluorescent molecules and biological molecules.

2D materials can form van der Waals (vdW) heterostructures with tailored electric, optoelectronic, and magnetic properties and functionalities.[29-31] For example, $MoS_2$/$WSe_2$ vdW heterostructures with a dual-gate architecture can be gate modulated to behave as either an Esaki diode with negative differential resistance or a backward diode with large reverse bias tunneling current.[32] However, the application of vdW heterostructures on SERS is still limited. Tan *et al.* reported that a graphene/$WSe_2$ heterostructure could greatly enhance the Raman signal of copper phthalocyanine (CuPc) *via* CM, and the stacking sequence determined the enhancement factor because electron transition probability rates were influenced by the different interlayer couplings.[33] Ghopry *et al.* reported that TMD nanodomes/graphene vdW structures could achieve a SERS sensitivity of rhodamine 6G (R6G) up to $10^{-12}$ M with the combination of CM and EM.[34]

In this work, we propose and demonstrate the use of vdW heterostructures of atomically thin BN and graphene (BN/G) to veil plasmonic nanoparticles for improved SERS. The top BN plays important roles in capturing probe molecules (especially aromatic molecules) and protecting the underneath graphene and Ag nanoparticles from oxidation for reusability. Due to the atomic thickness of the BN separation, probe molecules and graphene can have charge transfer for fluorescence quenching and Raman enhancement *via* CM. These synergistic effects enable a highly sensitive, reproducible, and reusable SERS substrate with a suppressed fluorescence background.

EXPERIMENTAL SECTION

**Preparation of BN/G heterostructures.** Atomically thin BN was mechanically exfoliated on silicon wafers covered by 90 nm silicon oxide ($SiO_2$/Si) from high-quality single-crystal hBN using Scotch tape.[35-36] An Olympus BX51 optical microscope equipped with a DP71 camera was used to locate monolayer (1L) BN, whose thickness was then confirmed by a Cypher atomic force microscope (AFM, Asylum Research). Graphene was prepared by following the same procedure, and its thickness was confirmed by Raman spectroscopy. Next, a ~50 nm thick polymethyl methacrylate (PMMA) film was spin-coated on the 1L BN. That is, three droplets of 2% PMMA (molecular weight of 495K) anisole solution were spin coated at 2000 rpm for 2 mins. After peeled off from the $SiO_2$/Si in 1 M NaOH aqueous solution, the PMMA/1L BN was transferred onto the top of the graphene under an optical microscope. The PMMA was then removed by acetone and acetic acid (Figure 1).

**Fabrication of BN/G heterostructure-veiled Ag nanoparticle SERS substrates (BN/G/Ag).**

A layer of Ag film (∼10 nm) was deposited on a $SiO_2$/Si wafer using a metal sputter (EM ACE600, Leica) with a sputtering current of 40 mA. The BN/G heterostructure was transferred on the top of the Ag film following the similar PMMA method, and then the PMMA was removed by acetone. Subsequently, the substrate was annealed in argon (Ar) at 450 °C for 1 h and cooled down to room temperature naturally (Figure 1). To form a close-to monolayer of R6G molecules, a droplet (20μL) of R6G ethanol solution ($10^{-6}$ M) was drop casted on each 15×15 mm substrates. Once the droplet was in contact with the substrates, it spread to the entire substrate almost instantly. With the help of a fan, the ethanol evaporated within 40 s. For reusability tests, the SERS substrates were heated in air at 360 °C for 5 min to remove R6G molecules. The Raman spectra of BN nanosheets, graphene and R6G were collected using a Renishaw inVia Raman microscope with a

514.5 nm laser. All Raman spectra were calibrated with the Raman band of Si at 520.5 cm$^{-1}$. An objective lens of 100× with a numerical aperture of 0.9 was used.

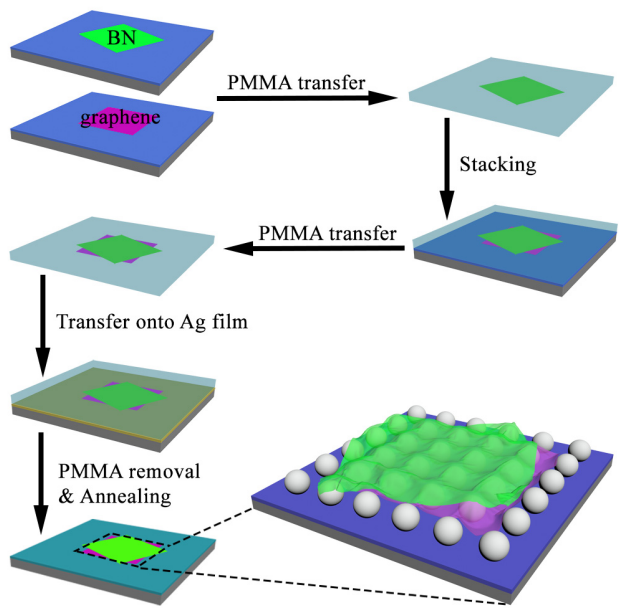

**Figure 1.** Schematic diagram of the preparation of BN/G/Ag substrate.

## RESULTS AND DISCUSSION

Figure 2a and b show the optical and AFM images of several pieces of 1L BN with thickness of

~0.61 nm on SiO$_2$/Si. Some graphene samples on SiO$_2$/Si are shown in Figure 2c, whose thickness

was confirmed by Raman spectroscopy (Figure 2d). Raman showed two sharp peaks: one at 1583.9

cm$^{-1}$ corresponding to the in-plane $sp^2$ C-C stretching mode with $E_{2g}$ irreducible representation (*G*

band), and the other at 2683.6 cm$^{-1}$ derived from the second-order Raman band from the in-plane

breathing-like mode of the carbon rings (*2D* band).[37] Both the low intensity (I) ratio of the *G* to

*2D* bands (I$_G$/I$_{2D}$ <1/2) and the symmetry of the *2D* band indicate the graphene was 1L.[38] No disorder-induced *D* band at ~1350 cm⁻¹ was observed, verifying the high quality and defect-free nature of the graphene samples.

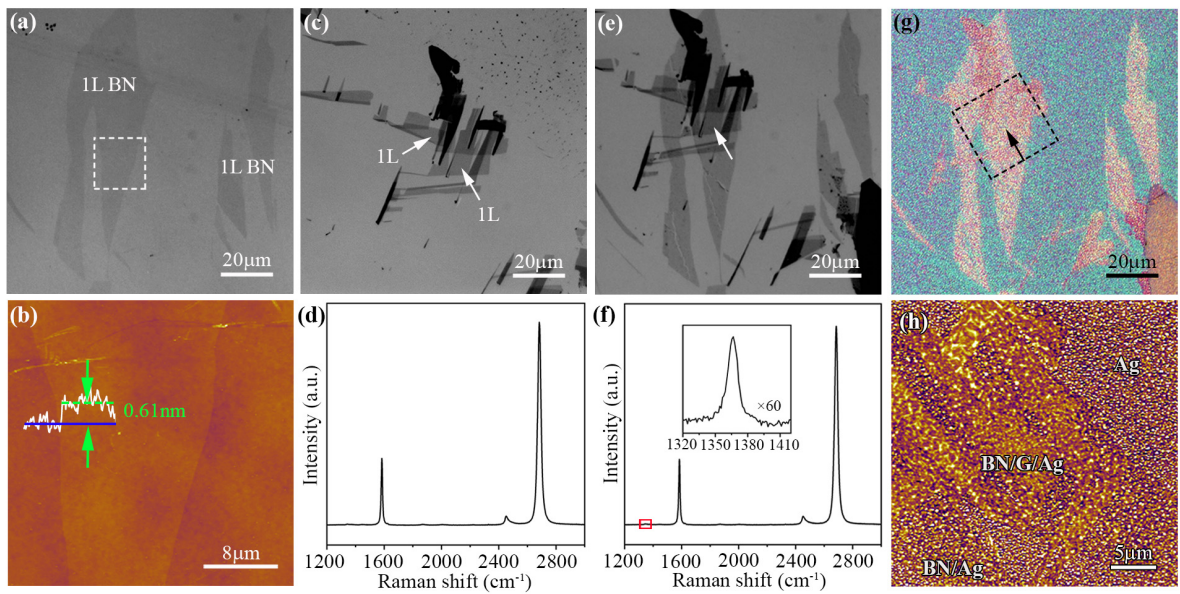

**Figure 2.** Optical (a) and AFM (b) images of as-exfoliated 1L BN samples on SiO₂/Si; optical image (c) and Raman spectrum (d) of as-exfoliated graphene of different thickness on SiO₂/Si; optical image (e) and Raman spectrum (f) of BN/G heterostructures after transfer, with an enlarged view of the Raman *G* band of the 1L BN in the red rectangle inserted; optical (g) and AFM (h) images of the SERS substrate of the BN/G heterostructures on Ag nanoparticles.

The BN/G heterostructure was fabricated by the PMMA-assisted transfer method (see details in the Experimental section). Figure 2e shows the optical image of the BN/G on SiO₂/Si substrate, where the 1L BN covered most of the graphene, including the two 1L graphene regions. Raman spectroscopy was used to characterize the BN/G heterostructure. The transferred graphene was intact, as revealed by the similar Raman spectrum with the same sharp *G* and *2D* bands at 1584.5 and 2684.7 cm$^{-1}$ as well as the absence of *D* band at ~1350 cm$^{-1}$ (Figure 2f). The 1L BN showed a weak *G* band at 1367.2 cm$^{-1}$ (insert in Figure 2f), confirming the successful fabrication of the desired vdW heterostructure. The BN/G heterostructures were then transferred by PMMA onto a 10 nm-thick Ag film on SiO$_2$/Si. After the PMMA cleaned off, annealing at 450 °C in Ar for 1 hour converted the Ag film to plasmonic nanoparticles, and BN/G/Ag/SiO$_2$/Si structures formed. Figure 2g displays the optical image of the BN/G/Ag SERS substrate. According to AFM, the BN/G followed closely the profile of underlying Ag nanoparticles, whose average size and distance were about 45 nm and 10 nm, respectively (Figure 2h and Figure S1). Such a structure can preserve plasmonic "hot spots" and effectively protect Ag nanoparticles from oxidation.[18,26]

The performance, including sensitivity/EF, fluorescence background quenching, and reusability of the BN/G/Ag SERS substrate was evaluated and compared with that of Ag nanoparticles covered by graphene or BN alone (*i.e.* G/Ag and BN/Ag) and without any coverage (Ag). One droplet (~20 μL) of 10$^{-6}$ M R6G solution was used as the probe molecules. EF was defined as ($I_{SERS}/C_{SERS}$)/($I_0/C_0$), where $I_0$ is the Raman intensity of 10$^{-3}$ M R6G on SiO$_2$/Si substrate without Ag nanoparticles (10$^{-3}$ M instead of 10$^{-6}$ M R6G was used because the signal of 10$^{-6}$ M was too weak without plasmonic enhancement), and *C* is the concentration of R6G. In order to accurately reflect the enhancement effect from the different substrates, about 20 Raman spectra from different areas of each substrate were collected (See Supporting Information, Figure S2). Figure 3a compares the typical Raman spectra of R6G from the different SERS substrates. The Ag substrate showed a strong fluorescence background (blue shaded area), reducing the SERS sensitivity. The fluorescence occurred when the excited electrons of R6G relax to the ground state. Specifically, the π electrons in carbon atoms (2$p_z$ orbits) and *n* electrons in oxygen atoms (nonbonding electrons) contained in the xanthene structure were excited by the laser to unoccupied excited energy states (π*), and then these π* electrons relaxed to the ground states of R6G by emitting photons.[39] The SPR of the Ag nanoparticles enhanced both the Raman and fluorescence signals of the R6G molecules close-to the hot spots *via* EM. Although the fluorescence from the R6G molecules directly contacting the Ag nanoparticles could be quenched because of the relaxation and transfer of the π* electrons to the Ag nanoparticles, most other R6G molecules located between Ag nanoparticles were hardly quenched, leading to the strong fluorescence background.

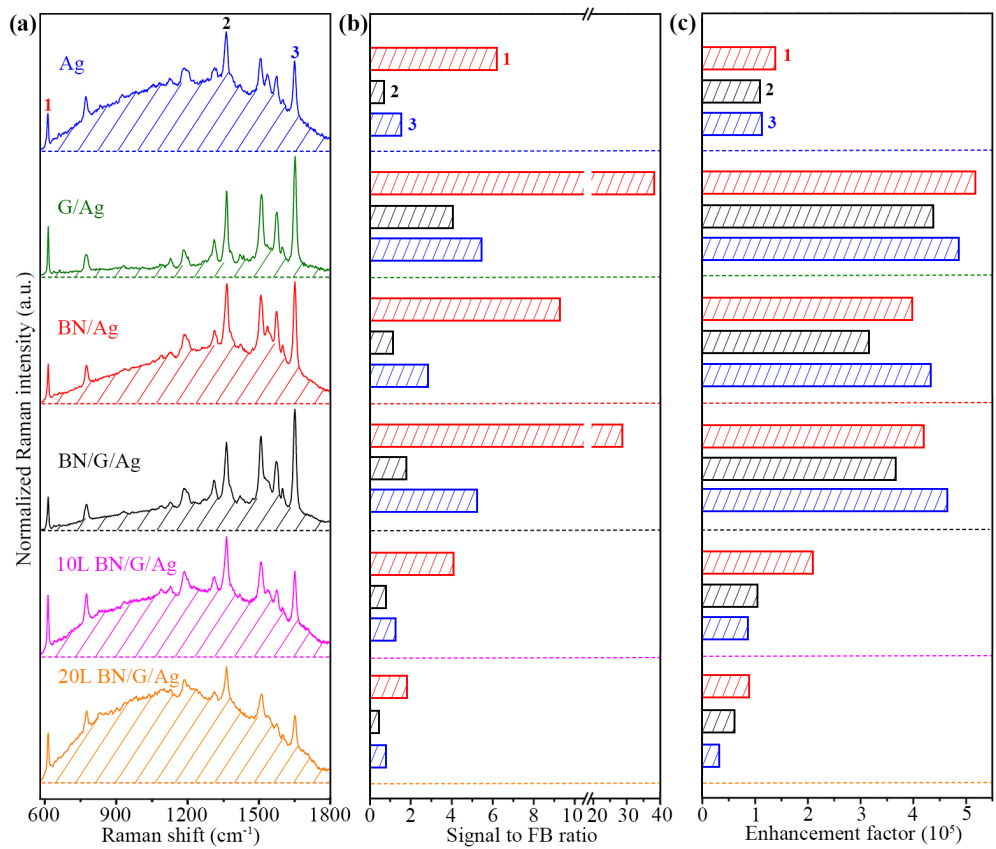

**Figure 3.** (a) Raman spectra of $10^{-6}$ M R6G on Ag (nanoparticles), (1L)G/Ag, (1L)BN/Ag, (1L)BN/(1L)G/Ag, 10L BN/(1L)G/Ag and 20L BN/(1L)G/Ag substrates; (b), (c) comparison of the signal-to-fluorescence background (FB) ratio and EF of the 613, 1363, and 1650 cm$^{-1}$ peaks of R6G from these SERS substrates.

In contrast, the G/Ag substrrate showed a much weaker fluorescence background (green in Figure 3a). The interaction between graphene and R6G enabled charge transfer that not only increased the Raman cross-section *via* CM and hence increased the SERS enhancement but also quenched the fluorescence background *via* the relaxation of R6G's π* electrons to graphene's π* band, followed by vibrational relaxation. Similar to what we reported before,[18] BN/Ag substrate could also greatly enhance the Raman signal of R6G, but a detectable fluorescence background was present (red in Figure 3a) mainly due to its electric insulation. Although R6G's π* electrons could tunnel through the 1L BN and charge transfer to the Ag nanoparticles,[40] it was impossible for the R6G molecules located between Ag nanoparticles to fulfill the charge transfer, similar to the case of the Ag substrate. In comparison, the BN/G/Ag substrate provided a strong Raman signal with much-suppressed fluorescence background (black in Figure 3a).

To illustrate the effect of the reduced fluorescence background on the Raman sensitivity, Figure 3b and c compare the signal-to-fluorescence background ratio and EF of the 613, 1363, and 1650 cm$^{-1}$ Raman peaks of R6G from the different SERS substrates. Take the 1363 cm$^{-1}$ peak as an example. The Ag substrate showed a S/FB ratio and EF of 0.67 and $1.11\times10^5$, respectively. The G/Ag substrate achieved the best S/FB (4.16) and EF ($4.40\times10^5$) due to the additional CM and reduced fluorescence background by graphene. The BN/Ag had a S/FB ratio of 1.10 and EF of $3.15\times10^5$. The BN/G/Ag substrate enhanced the S/FB by 168% and 64% and EF by 229% and 16% in comparison with those of Ag and BN/Ag substrates, respectively. The improved EF mainly came from the lower fluorescence background, which disguised the Raman signal and therefore deteriorated the Raman sensitivity. The following reasons contributed to these improvements: 1) electrons could tunnel through 1L BN, and the charge transfer between R6G molecules and graphene led to fluorescence quenching and the CM in addition to EM; 2) the high flexibility of BN/G made the heterostructure closely follow the profile of underlying Ag nanoparticles, preserving the hot spots to the maximum extent. On the other hand, atomically thin BN has an excellent absorption capability towards aromatic molecules due to π-π interaction and conformation change, which can further increase Raman sensitivity in practical applications.[27] We also tested another analyte, i.e. methylene blue, and similar SERS improvements were obtained from the BN/G/Ag substrate, demonstrating the universality of the method (See Supporting Information, Figure S3).

To justify the above mechanisms, we also prepared BN/G/Ag SERS substrates using thicker (10L and 20L) BN, namely 10LBN/G/Ag and 20LBN/G/Ag. Their SERS performances are also included in Figure 3. The fluorescence background increased obviously with the incremental thickness of BN (pink and orange in Figure 3a), while the Raman enhancements decreased (Figure 3b and c). This was attributed to 1) the SPR of Ag nanoparticles decayed exponentially with increased distance from the hot spots, and 2) worse electron tunneling effect on the thicker BN, leading to stronger fluorescence background and hence lower sensitivity.[40-41]

Atomically thin BN can effectively protect metals from oxidation due to its excellent thermal stability and gas impermeability.[28,42] Ag nanoparticles veiled by BN have a long "shelf-time" due to the lack of oxidation-induced sensitivity loss. More importantly, BN-protected SERS substrates can be regenerated by heating in air to burn off probe molecules and then reused.[16,18,26] We tested the reusability of the BN/G/Ag substrate and compared the result with the BN/Ag, G/Ag and Ag substrates (Figure 4). In each reusability cycle, ~20 μL R6G solution ($10^{-6}$ M) was dropped on the four substrates and removed by heating at 360 °C in the air for regeneration. Heating in air has been reported the most effective and efficient way to remove probe molecules for reusable SERS.[16] Similar to our previous studies, Raman intensity of R6G from both BN/G/Ag and BN/Ag substrates remained essentially unchanged after 5 cycles of reusability tests (Figure 4a-b), implying that the Ag nanoparticles were protected against oxidation even after repeated heat treatments in air owing to the BN coverage.

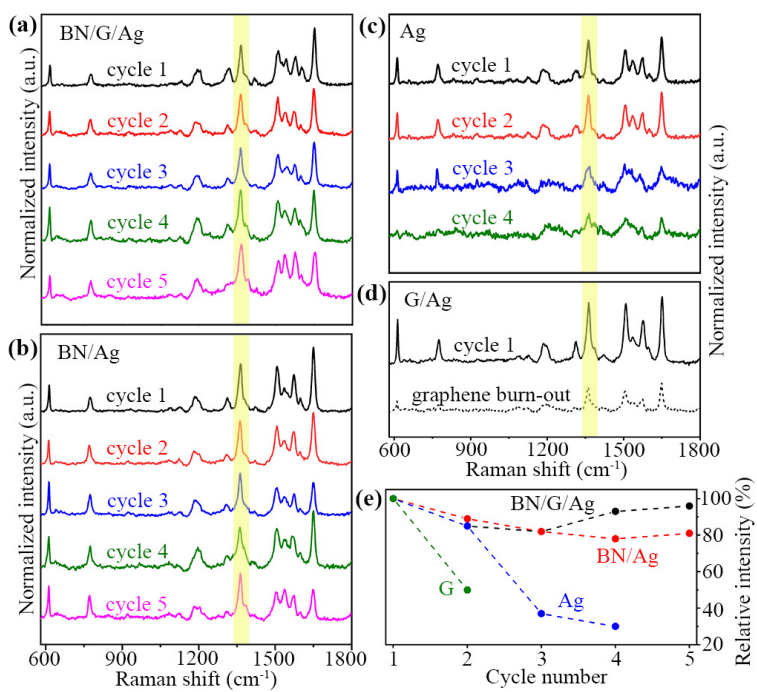

**Figure 4.** Reusability tests of BN/G/Ag (a), BN/Ag (b), Ag (c) and G/Ag (d) substrates; (e) the relative intensity change of the 1363 cm⁻¹ Raman peak of R6G (highlighted in a-d) from the four substrates.

In contrast, the SERS intensity of R6G from the Ag substrate decreased dramatically after four times of reuse (Figure 4c), indicating severe oxidation of the Ag nanoparticles without the protection of BN. For G/Ag substrate, the signal intensity of R6G halved just after one cycle of reuse (Figure 4d) due to the vanish of graphene after heat treatment in air.[18] The reusability of the different substrates was quantified by the intensity change of the 1363 cm⁻¹ Raman peak of R6G, as summarized in Figure 4e. Both BN/G/Ag and BN/Ag substrates kept more than 80% of their plasmon enhancement after five times of reusability tests (black and red). However, the Ag substrate lost ~70% of its intensity after four cycles of reusability test because of oxidation, and G/Ag substrate was not reusable at all due to the low thermal stability of graphene, which is discussed in details below.

To confirm the protection by the 1L BN, the Raman spectra of graphene from BN/G/Ag and G/Ag substrates after the reusability tests were studied. After five times of reusability test, the Raman signal of graphene in BN/G/Ag was still strong: a symmetric *2D* band and $I_G$/$I_{2D}$ of ~0.5. It confirmed the existence of graphene after several times of heat treatments at 360 °C in air, as shown in Figure 5 (black). However, a *D* band at 1351.8 cm⁻¹ appeared, indicating slight oxidation of the graphene. According to $L_D^2 (nm^2) = (1.8 \pm 0.5) \times 10^{-9} \lambda_L^4 (I_D/I_G)^{\wedge}(-1)$, where $\lambda_L$ is excitation laser wavelength in nanometers, one can calculate the distance between defects, $L_D$. The ratio of $I_D/I_G$ was about 0.49, and hence $L_D$=16 nm, suggesting a relatively low defect density in the graphene. As high-quality BN sheets are gas impermeable,[28,42] $O_2$ molecules cannot pass through the BN to oxidize the graphene. It was likely that a small number of oxygen molecules entered the gaps of the Ag nanoparticles and oxidized the graphene during the heat treatments (as shown in Figure 1). In contrast, no Raman signal of graphene was detected from the G/Ag substrate just after one cycle of regeneration (green in Figure 5). We note that the Gibbs free energy of the oxidation of graphene at 360 °C is about −400 KJ/mol, indicating that graphene reacts with oxygen spontaneously at such high temperature without the protection of BN nanosheets. The burn-out or disappearance of graphene led to the oxidation of the Ag nanoparticles and the loss of fluorescence quenching, surface absorption capability, and CM enhancement. As a result, the G/Ag substrate lost >50% of its Raman enhancement just after one cycle of the reusability test and was not reusable at all.

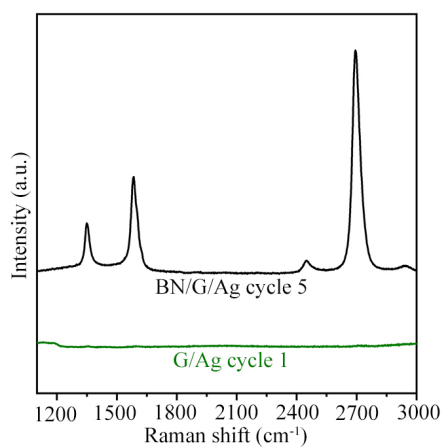

**Figure 5.** Raman spectra of graphene from BN/G/Ag (black) and G/Ag substrates (green) after the reusability tests.

CONCLUSIONS

In summary, the vdW heterostructures of atomically thin BN and graphene were demonstrated as a multi-functional coverage of plasmonic Ag nanoparticles for SERS. It took advantage of both graphene and BN. Although probe molecules and graphene were separated by a monolayer BN, electrons could tunnel through the atomically thin BN to allow charge transfer between probe molecules and zero-bandgap graphene, leading to the enhancement of Raman signals and fluorescence background quenching. The atomically thin BN had high thermal stability and excellent impermeability to protect the underlying graphene and Ag nanoparticles from oxidation even at high temperatures in air, enabling the regeneration of the SERS substrate by heat treatment at 360 °C. As a result, a highly sensitive, fluorescence background-suppressed, and reusable SERS substrate was achieved. These advances will broaden the applications of SERS in various fields, *e.g.* biological imaging.

AUTHOR INFORMATION

**Corresponding Author**

*Email: luhua.li@deakin.edu.au

**Notes**

The authors declare no competing financial interests.

ACKNOWLEDGEMENTS

Q.C. acknowledges Deakin University under Alfred Deakin Postdoctoral Research Fellowship 2018. L.H.L thanks the financial support from Australian Research Council (ARC) *via* Discovery

Early Career Researcher Award (DE160100796). K.W. and T.T. acknowledge the support from the Elemental Strategy Initiative conducted by the MEXT, Japan and the CREST (JPMJCR15F3), JST. S.H. acknowledges Chinese NSFC funding 51920105004.